\documentclass[12pt,preprint]{aastex}

\def\Msunh{\mbox{$h^{-1}\ M_\odot$}}

\shorttitle{Mass Function High Mass Behaviour}
\shortauthors{Betancort-Rijo \& Montero-Dorta}

\begin{document}

\title{Understanding the Cosmic Mass Function High Mass Behaviour}

\author{Juan E. Betancort-Rijo$^{1,2}$, Antonio D. Montero-Dorta$^{2,1}$}

\affil{$^{1}$Instituto de Astrof\'{i}sica de Canarias, V\'{i}a L\'{a}ctea s/n, 
 La Laguna, Tenerife, E38200, Spain}
\affil{$^{2}$Universidad de La Laguna, Departamento de Astrof\'{i}sica, Avda. Astrof\'{i}sico Fco. Sánchez s/n. ,E38200, La Laguna, Tenerife, Spain}
 
\email{jbetanco@iac.es}

\begin{abstract}
We claim that the discrepancy found between theoretical predictions for the cosmic mass function and those found in numerical simulations is due to the fact that in deriving the former all mass elements are assumed to be at the center of the object they belong to (the all-mass-at-center problem). By an appropriate treatment of this problem, using the spherical collapse model (which is not a bad approximation in the high mass limit), we obtain both the high mass behaviour found in simulations and the true assymptotic behaviour of the mass function (for arbitrarily high masses). Therefore, we conclude that by combining ellipsoidal dynamics with a suitable treatment of the all-mass-at-center problem (that we will show in a follow up work) a theoretical prediction for the cosmic mass function in full agreement with simulations may be obtained.    
\end{abstract}


\section{Introduction}
Cosmic mass functions corresponding to various cosmological models and collapsed object defining criterias have been accurately determined (\citet{Jen01}, \citet{War06}) by means of numerical simulations covering five orders of magnitude in mass. On the other hand, theoretical ideas concerning these functions have developed considerably from those contained in a seminal work by \citet{PS74}. In this work, the simple dynamics of the spherical collapse model was combined with Gaussian statistics to derive an expression for the cosmic mass function. However, one of the criticism that can be made to this derivation is that mass elements lying at the center of underdense (or not sufficiently overdense) density fluctuations on a certain scale are assigned to collapsed objects smaller than this scale, neglecting the possibility that these elements could lie at the center of a sufficiently overdense fluctuation on a larger scale (so that it should be assigned to a collapsed object larger than this latter scale). To address this difficulty, currently referred to as the cloud-in-cloud problem, \citet{PH90} and \citet{Bo91}, using for Gaussian fields ideas advanced by \citet{Ep83} for Poissonian point distributions, developed the, so called, excursion set approach, whereby every mass element is assigned to an object with the mass corresponding to the largest sphere centered at this element and with mean inner overdensity such that accordingly with the spherical collapse model it has just virialized. \\
Even with this last refinements, theoretical predictions for the mass function did not provide a perfect fit to the results found in numerical simulations (\citet{GB94}). This prompted questioning the suitability of the spherical collapse model (\citet{BS81}). To deal with this issue \citet{BM96} used ellipsoidal dynamics within a general framework for identifying in the initial conditions the points at which objects of different masses will form, and following their motion. \citet{SMT01} discussed a simple procedure for incorporating ellipsoidal dynamics into the excursion set formalism, and developed it to fuller extent in \citet{ST02}. Their prediction for the mass function is in much better agreement with those found in numerical simulations, specially in the low mass limit where it seems to give the correct logarithmic slope. However, in the high mass limit, where the functional behaviour of their mass function is dominated by the same exponential factor that dominates Press-Schechter expression, namely:
 
 \begin{equation}
\exp\bigg(-a\frac{\delta_{c}^{2}}{2\sigma^{2}(m)}\bigg)
\label{eq:eq1}
\end{equation}  
 
with $a$ equal to $1$ ($\sigma(m)$ being the rms linear density fluctuation as a function of mass), the discrepancy is larger than a factor $10$ for the largest masses available ( $\sim 5\cdot 10^{15} \Msunh$ ). A value of $a$ around $0.75$ in factor (\ref{eq:eq1}) with $\delta_c\sim1.7$ is in much better agreement with numerical results.\\
This discrepancy poses a rather vexatious question which is not easy to justify theoretically. If collapsed objects are defined as spherical overdensities with density contrast $\Delta$ and $\delta_c$ is its linear counterpart through the spherical collapse model (including shell-crossing non-radial motions and radial velocity dispersion), \citet{SMT01} procedure leads inescapably to (\ref{eq:eq1}) with $a=1$ for the leading factor of the high mass limit. This is so because for the largest masses the collapse is almost spherical and the expression obtained using the spherical collapse model must be recovered.\\
\citet{ST02} derivation of the mass function has certain amount of uncertainty; contained in the way in which they modelled the process of virialization of the proto-objects, where massive shell-crossing takes place. However, this model has been built so that for spherically collapsing patches the actual density contrast, $\delta$, takes the value $\Delta$ when the linear density contrast , $\delta_l$, takes the value $\delta_c$. So, the above result is not affected by the mentioned uncertainty.\\
For $\Delta \sim 200$ the value of $\delta_c$ is around $1.7$ ($1.686$ is the conventional value for Einstein-deSitter cosmology). In this case, as mentioned above, by comparing theoretical predictions with simulations one obtains $a \sim 0.75$. However, even if we were to use the standard spherical model (neglecting the dynamical ralentization due to shell-crossing and other factors), $\delta_c$ must be close to $1.6$, so that $a$ must be around $0.85$ to agree with simulations. Now, any refinement in the dynamical model used (i.e. including substructure) leads to a larger value of  $\delta_{c}$,therefore, the fact that $a$ is close to $0.75$, cannot be explained in this manner. So, we must search for its explanation in another assumption of the formalism.\\
In this work we show that the failure of theoretical predictions to render the correct value of $a$ is due to the fact that in the excursion set formalism (as well as in that of Press-Schechter) the criteria for deciding to objects of which mass must be assigned a mass element is based upon the global behaviour of patches (spherical or ellipsoidal) centered at that element (all-mass-at-center). Obviously, of all mass elements making up an object only one is exactly at the center, but assuming for simplicity every element to be at the center of the objects it belongs to is not, generally, a bad approximation. However, as we shall show, for large masses it leads to an underestimation of the mass function.\\
We show here that if, instead of assuming every mass element to be at the center of its object (with mass m), we merely assume that it is closer than $Q(m)$ ($Q$ being the Lagrangian radius of the object)  to the real central element, the value of $a \sim 0.75$ comes out nicely.\\
It must be noted that in the algorithm of \citet{BM96} for obtaining mass functions from realizations of initial conditions, the problem is implicitly treated like here, but not explicit analytical treatment has so far been available. Furthermore, the question of all-mass-at-center has not been investigated in regard with the present problem. If Bond \& Myers algorithm were used to this purpose, the results shown here would also be found.\\
In this work we only consider spherical dynamics, since the effect considered is only important for large masses where the collapse is almost spherical. This simpler dynamics helps making more patent the essence of the effect, which shall be treated to full extent using triaxial collapse in a follow up paper.

\section{Dealing with the all-mass-at-center problem}

Let $F(m)$ be the fraction of mass making up objects with mass larger than m, and n(m) the number density of objects within a unit mass interval around m. The following identity must obviously hold:

\begin{equation}
n(m)=-\frac{\rho_b}{m}\frac{dF(m)}{dm}\label{eq:eq2}
\end{equation}

This relationship, in various forms (using Ln(m), $\sigma$(m), Ln($\sigma$(m)) as independent variable), is used in most theoretical derivations of the mass function, in particular, in the excursion set and Press-Schechter formalisms. It is the approximations used to obtain F(m) that make the difference between various approaches. For example, the dynamics can be modelled by the spherical collapse model or by the more refined ellipsoidal model. Here we adopt the spherical model since the effect we intend to explain, namely, the "anomalous" large mass behaviour of the mass function does not depend significantly on the dynamical model used.\\
In the excursion set approach with spherical dynamics F(m) is equated to the probability that for a randomly chosen mass element the largest sphere centered at this element with an inner linear density fluctuation equal to $\delta_{c}$ has a radius larger than $Q(m)$. That is, as mentioned in the introduction, the computation of n(m) proceeds as if all mass elements were at the center of the object they make up. This assumption is common to all approaches using expression (\ref{eq:eq2}) developed so far. The peak approach (\citet{PH90}) does not make this assumption but it has other problems that we shall discuss in a follow up work.\\
The aim of this work is to show that an approach using expression (\ref{eq:eq2}) may easily be developed where F(m) is computed applying the spherical collapse model only to spheres centered at the central point of each object, and then show that the resulting mass function has the correct large mass behaviour.\\
If  the spherical collapse model applies, a randomly chosen mass element belonging to an object of mass m must lie, in the initial conditions, within a distance (lagrangian) Q(m) from a point such that a sphere of radius Q(m) centered at it has an enclosed mean linear density contrast, $\delta_l$, equal to $\delta_c$.\\
Let us now obtain the probability distribution for $\delta_l$ within a sphere of radius Q(m) centered, not at the center of the proto-object but at a randomly chosen point within it. When such point is at a distance (Lagrangian) q from the center,  the Gaussian bivariate distribution for the linear fields in scale Q(m) at q and at $q=0$ (the center) leads straightforwardly (\citet{Pa06}, expression (24)) to the following expression for the distribution of $\delta_l$ at q,\quad|$\delta_{l}(q)$|\quad, conditioned to $\delta_{l}(q=0)=\delta_c$:

\begin{equation}
P(\delta_{l}(q)|\delta_{l}(q=0)=\delta_c,Q)=\frac{1}{\sqrt{2\pi}}\frac{ \exp\bigg(\frac{-(\delta_{l}(q)-\delta_{c}S(q))^2}{2\sigma\prime(q)^{2}}\bigg)}{\sigma\prime(q)}  \label{eq:eq3}
\end{equation}

\begin{displaymath}
\sigma\prime(q)=\sigma(m)(1-S(q)^{2})^{\frac{1}{2}};\quad S(q)\equiv S(q,Q)\equiv\frac{\sigma_{12}(q,Q)}{\sigma(m)^{2}}
\end{displaymath}

\begin{displaymath}
\sigma_{12}(q,Q)=<\delta_{l}(q)\delta_{l}(q=0)>=\frac{1}{2\pi^{2}q}\int_{0}^{\infty}\mid\delta_{k}\mid^{2}W(Qk)^{2}sin(kq)kdk
\end{displaymath}

\begin{displaymath}
W(x)=\frac{3}{x^{3}}(sin~x~-x~cos~x~); \quad        Q \equiv Q(m)
\end{displaymath}

where $\mid\delta_{k}\mid^{2}$ stands for the linear power spectra. $S(q,Q)$ may accurately be approximated by:

\begin{equation}
S(q,Q)\simeq e^{-c(\frac{q}{Q})^{2}}; c\equiv c(m)\equiv -Ln(S(Q,Q))  \label{eq:eq4}
\end{equation}

To obtain the conditional probability distribution ,$P_{1}(\delta_l|\delta_l(q=0)=\delta_c,Q)$, for $\delta_l$ (on scale Q(m)) at a randomly chosen point within the proto-object given that $\delta_l(q=0)=\delta_c$, we multiply expression (\ref{eq:eq3}) by the probability distribution for q at a randomly chosen point within $Q$ and integrate over q:

\begin{equation}
P_{1}(\delta_l|\delta_l(q=0)=\delta_c,Q)=\int_{0}^{Q}P(\delta_l(q)|\delta_l(q=0)=\delta_c,Q)\frac{3q^{2}}{Q^{3}}dq   \label{eq:eq5}
\end{equation}				

When $\delta_l(q=0)$, instead of taking the value $\delta_{c}$ takes a larger value, the resulting collapsed object (centered at $q=0$) would have a mass, $m\prime$, somewhat larger than m. So, the probability distribution, $P_{2}(\delta_{l}|m\prime>m)$, for $\delta_l$ (on scale $Q(m)$) for a point chosen at random within a proto-object with mass $m\prime$ larger than $m$ (that is, for a point lying at a distance smaller than $Q(m\prime)$ from the center of their proto-objects), may be obtained by multiplying the conditional distribution given in expression (\ref{eq:eq5}) (with $\delta_{l}(q=0)$ as variable in the place of  $\delta_c$ in (\ref{eq:eq3})) by the probability distribution for $\delta_{l}(q=0)$, integrating for $\delta_l(q=0)\geq\delta_c$, and dividing by the probability that $\delta_l(q=0)\geq\delta_c$:
              
\begin{equation}
P_{2}(\delta_l|m\prime>m)=\frac{\int_{\delta_c}^{\infty}\frac{\exp-\frac{1}{2}\frac{\delta_l^{2}(q=0)}{\sigma^{2}(m)}}{\sqrt{2\pi}\sigma(m)}P_{1}(\delta_l|\delta_l(q=0),Q)d(\delta_l(q=0))}{\frac{1}{2}erfc\big(\frac{\delta_c}{\sqrt{2}\sigma(m)}\big)}     \label{eq:eq6}
\end{equation}

It is clear that expression (\ref{eq:eq6}) cannot be exact, because in expression (\ref{eq:eq5}) $m\prime=m$ is assumed, but for $\delta_{l}(q=0)$ larger than $\delta_{c}$ $m\prime$ is somewhat larger than m. However using (\ref{eq:eq6}) leads only to a minute error that shall be discussed in a future work. Carrying out the integral over $\delta_{l}(q=0)$ in (\ref{eq:eq6}), we find:

\begin{equation}
P_{2}(\delta_l|m\prime>m)=\frac{\exp\big(-\frac{\delta_{l}^{2}}{2\sigma(m)^{2}}\big)3\int_{0}^{1}\frac{1}{2}erfc\big(\frac{\delta_{c}}{\sqrt{2}\sigma(m)}\big(\frac{1-\exp(-cu^{2})}{1+\exp(-cu^{2})}\big)^{\frac{1}{2}}\big)u^{2}du}{\sqrt{2\pi}\sigma(m)\frac{1}{2}erfc\big(\frac{\delta_{c}}{\sqrt{2}\sigma(m)}\big)}    \label{eq:eq7}
\end{equation}

where use has been made in expression (\ref{eq:eq3}) of the approximation given in (\ref{eq:eq4}) and where the variable q has been changed to u ($\equiv\frac{q}{Q}$).\\
When  $\delta_{l}$ is larger than $\delta_{c}$ for a sphere of radius $Q(m)$ centered at a randomly chosen mass element, it is clear that this element must belong to an object of mass larger than m. So, the probability distribution for $\delta_{l}$ within a sphere of radius $Q(m)$ centered at a randomly chosen mass element must be, for $\delta_{l}>\delta_{c}$, equal to the probability distribution for $\delta_{l}$ for elements chosen at random amongst those belonging to objects with mass larger than $m$ (i.e. $P_{2}(\delta_l|m\prime>m)$) multiplied by the probability that a randomly chosen element belongs to an object larger than $m$. But this later probability is simply the fraction of mass belonging to objects with mass larger than $m$, which we represent by $F(m)$. We may then write:

\begin{equation}
F(m)P_{2}(\delta_l|m\prime>m)=\frac{\exp\big(-\frac{\delta_{l}^{2}}{2\sigma(m)^{2}}\big)}{\sqrt{2\pi}\sigma(m)} \quad for \quad \delta_{l}>\delta_{c}
\label{eq:eq8}
\end{equation}

Using expresion (\ref{eq:eq7}) we find:

\begin{equation}
F(m)=\frac{\frac{1}{2}erfc\bigg(\frac{\delta_{c}}{\sqrt{2}\sigma(m)}\bigg)}{\frac{1}{2}V(m)}  \label{eq:eq9}
\end{equation}

\begin{displaymath}
V(m)\equiv3\int_{0}^{1}erfc\bigg(\frac{\delta_{c}}{\sqrt{2}\sigma(m)}\bigg(\frac{1-\exp(-c(m)u^{2})}{1+\exp(-c(m)u^{2})}\bigg)^{\frac{1}{2}}\bigg)u^{2}du
\end{displaymath}

From expression (\ref{eq:eq9}) we see that $\frac{V(m)}{2}$ is simply the ratio between the probability that $\delta_{l}\geq\delta_{c}$ within a sphere of radius $Q(m)$ centered at a randomly chosen mass element and the larger probability that a random mass element belongs to an object of mass larger than $m$.

\section{The Mass Function}

Using in expression (\ref{eq:eq2}) $F(m)$ as given by approximation (\ref{eq:eq9}) we find that the mass function, $n(m)$, may be written in the form:

\begin{equation}
n(m)=\bigg[-\bigg(\frac{2}{\pi}\bigg)^{\frac{1}{2}}\frac{\rho_{b}}{m}\frac{\delta_{c}}{\sigma(m)^{2}}\exp\bigg(-\frac{\delta_{c}^{2}}{2\sigma^{2}(m)}\bigg)\frac{d\sigma}{dm}\bigg]\frac{[1-Z(m)]}{V(m)}
\label{eq:eq10}
\end{equation}

\begin{displaymath}
Z(m)\equiv\bigg(\frac{\pi}{2}\bigg)^{\frac{1}{2}}\frac{\sigma(m)^{2}}{\delta_{c}}\exp\bigg(\frac{\delta_{c}^{2}}{2\sigma^{2}(m)}\bigg)erfc\bigg(\frac{\delta_{c}}{\sqrt{2}\sigma(m)}\bigg)\frac{dLnV(m)}{d\sigma(m)} 
\end{displaymath}

In figure (1) we show $c(m)$ (see expression (\ref{eq:eq4})), $V(m)$, $Z(m)$ for a CDM power spectrum (BBKS 1986) with $\sigma_{8}=0.9$, $h=0.7$,$\Omega_{m}=0.3$ and with $\delta_{c}=1.676$. It is apparent that $Z(m)$ is small compared with $1$ for all masses (its largest value is about 0.26). So, expression (\ref{eq:eq10}) is essentialy the normalized Press-Schechter mass function divided by V(m). Therefore, it is this function that makes the most difference between the full mass function given by the spherical collapse model and the approximations to this function provided by either Press-Schechter or excursion set formalisms. In the small mass limit V(m), (which is by definition smaller than one for any mass), goes to one: it is larger than $0.9$ for masses smaller than $10^{11} \Msunh$ and it is still larger than $\frac{2}{3}$ for $3\cdot10^{13}\Msunh$. It is only for larger masses that it becomes a steep function of mass, going down to $0.06$ for $5\cdot10^{15}\Msunh$. So, it is at the high mass limit ($m\gtrsim10^{14}\Msunh$) where neglecting the all-mass-at-center problem leads to substancial errors.\\

From expression (\ref{eq:eq10}) and figure (1) it is apparent that the difference with the normalized P-S mass function (first factor in (\ref{eq:eq10})) is considerable only for $m\gtrsim10^{14}\Msunh$, giving this expression a substantialy enhanced density for the largest masses that have collapsed at present. For smaller masses the difference between both mass functions is small. However, expression (\ref{eq:eq10}) must give smaller densities than P-S in the small mass limit to compensate for the high mass excess, since both mass functions are normalized. This is actually the case for masses smaller than roughly $10^{13}\Msunh$, the difference increasing slowly with decreasing mass and stabilizing around a $20\%$ for masses below $10^{10}\Msunh$. This difference is mostly due to $Z(m)$, ($Z(10^{8}\Msunh)=0.25$), since $V(m)$ is close to one for these masses.\\

The interest of the low mass limit is mostly academic, since ellipsoidal collapse plays an important role in this limit. In the high mass limit, however, the collapse is close to spherical, so that the ratio between expression (\ref{eq:eq10}) and P-S (i.e. the second factor in (\ref{eq:eq10})) should be basically the same as that between the actual mass function and the mass function obtained using ellipsoidal dynamics but with the excursion set approach. To support this conjecture, we show in figure (2) the ratio between expression (\ref{eq:eq10}) and P-S (normalized) along with the ratio between Warren et al's (2006) fit to numerical results and Sheth \& Tormen mass function ($a=1$) (\citet{SMT01}). The basic agreement between both ratios strongly suggest that the appropriate treatment of the all-mass-at-center problem is all that one needs to add to Sheth \& Tormen derivation to obtain a mass function in full agreement with simulations.\\
The ratio between expresion (\ref{eq:eq10}) and P-S, for the mass range considered, is essentialy given by $V(m)^{-1}$ (expression (\ref{eq:eq9})). We see, from figure (2), that for $m\lesssim3\cdot10^{15}\Msunh$, $V(m)$ may be approximated by a function of the form given in expression (\ref{eq:eq1}) with $a\simeq0.25$.\\  
So, in the high mass limit, expression (\ref{eq:eq10}) is dominated by a term of the form (\ref{eq:eq1}) with $a \simeq 0.75$. The fact that $a$ is smaller than one, that we called "anomalous assymptotic behaviour", is therefore explained by the approximately exponential behaviour of $V(m)$.\\
It must be noticed, however, that $V(m)$ cannot take the form given in (\ref{eq:eq1}) for very large masses. In fact, it may easily be shown that in the arbitrarily high mass limit (as opposed to the limit of the highest masses available, that we have so far considered) the dependence of V on $\sigma(m)$ is not exponential, but potential:

\begin{equation}
V(m)\sim\ 0.564\bigg(\frac{c(m)^{\frac{1}{2}}}{2}\frac{\delta_{c}}{\sigma(m)}\bigg)^{-3}  (m\rightarrow\infty)   
\label{eq:eq11}
\end{equation}

So, in the truly assymptotic limit, expression (\ref{eq:eq10}) behaves as expression (\ref{eq:eq1}) with $a=1$, which, as we mentioned in the introduction, is the value that $a$ must take in this limit. Using (\ref{eq:eq11}) in (\ref{eq:eq10}) leads to the same assymptotic expression as using, in this limit, the peak approach, as we shall show in detail in a forthcoming paper.\\
Summarizing, the high mass behaviour of the cosmic mass function observed in numerical simulations, although dominated by a factor of the form given in (\ref{eq:eq1}), do not yet shows the real assymptotic behaviour. The apparent assymptotic behaviour led by expression (\ref{eq:eq1}) with $a\simeq0.75$ results from the approximate exponential behaviour of $V(m)$ in the relevant range of masses. The real assymptotic behaviour of $V(m)$ is potential with respect to $\sigma(m)$, resulting in a true assymptotic behaviour for $n(m)$ which takes the form (\ref{eq:eq1}) with $a=1$; as it must be. The function $V(m)$, which makes the essential difference with previous approaches, originates from an appropriate treatment of the all-mass-at-center problem.\\

\clearpage

\begin{figure}
\includegraphics[width=3cm,height=8cm,angle=90]{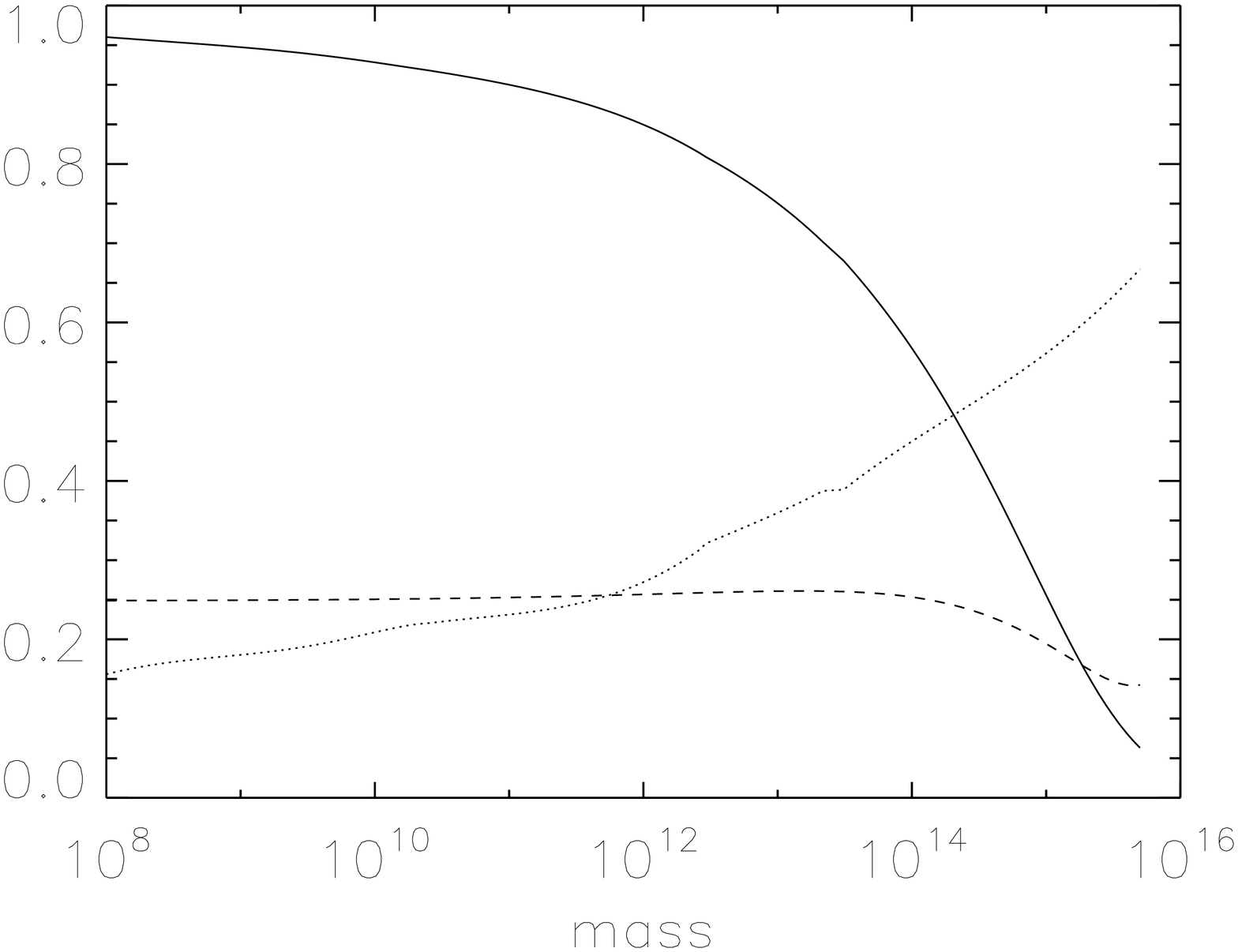}
\caption{{\small\small $c(m)$ defined in (\ref{eq:eq4}) is given by the dotted line; $V(m)$ (defined in (\ref{eq:eq9}) is given by the solid line; and $Z(m)$ (expression (\ref{eq:eq10})) is given by the dashed line.}}
\end{figure}

\clearpage

\begin{figure}
\includegraphics[width=3.9cm,height=8cm,angle=90]{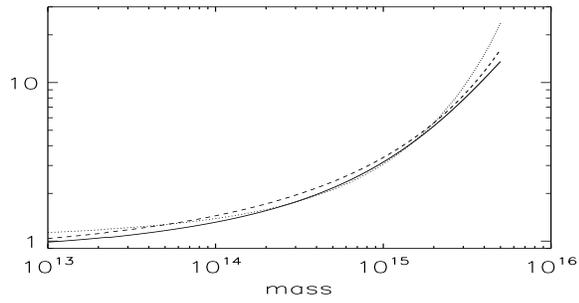}
\caption{{\small Ratio between the mass function as given by expression (\ref{eq:eq10}) and as given by the normalized Press-Schechter expression (solid line); an approximation to this ratio given by expression (\ref{eq:eq1}) with $a=0.25$ (dotted line); and the ratio between Warren et al's (2006) fit to numerical results and Sheth \& Tormen mass function ($a=1$) (dashed line). All theoretical expressions use $\delta_{c}=1.676$.}}
\end{figure}

\end{document}